\documentclass[aps,pre,reprint,superscriptaddress]{revtex4-1}
\usepackage{amsmath,graphicx,color,epsfig}
\usepackage{graphicx,color,epsfig}
\usepackage{amssymb,amsmath,enumerate}

\begin{document}
\title{Mass media competition and alternative ordering in social dynamics}
\author{O. Alvarez-Llamoza}
\affiliation{Universidad Cat\'olica de Cuenca, C²MAD-CIITT, Cuenca 010101, Ecuador.}
\author{M. G. Cosenza}
\affiliation{School of Physical Sciences and Nanotechnology, Universidad Yachay Tech, Urcuqu\'i, Ecuador.}
 \author{J. C. Gonzalez-Avella}
\affiliation{Institute for Cross-Disciplinary Physics and Complex Systems, UIB-CSIC,  Palma de Mallorca, Spain.}
\affiliation{Advanced Programming Solutions SL, Palma de Mallorca, Spain.}
\author{M. A. Suarez}
\affiliation{School of Physical Sciences and Nanotechnology,  
Universidad Yachay Tech, Urcuqu\'i, Ecuador.}
\author{K. Tucci}
\affiliation{SUMA-CeSiMo, Universidad de Los Andes, M\'erida, Venezuela.}
\author{P. Valverde}
\affiliation{Pontificia Universidad Cat\'olica del Ecuador, Facultad de Ciencias Exactas y Naturales, Quito, Ecuador.}

\begin{abstract}
We investigate the collective behavior of a system of social agents subject to the competition between two mass media influences considered as external fields. We study under what conditions either of two mass media with different intensities can impose its message to the majority. 
In addition to a collective state dominated by the stronger mass media and
a disordered phase, we characterize two nontrivial effects as the parameters of the system are varied: (i) the appearance of a majority sharing the state of the weaker mass media, and (ii) the emergence of an alternative ordering in a state different from those of either media.
We explore the dependence of both phenomena on the topology of the network of interactions. We show that the presence of long-range interactions rather than random connections is essential for the occurrence of both effects.
The model can be extended to include multiple mass media and we illustrate it by considering
three mass media fields acting on the system. Nontrivial collective behaviors persist for some ranges of parameters: the weakest mass media can convince the majority, and the system
can spontaneously order against all applied fields.

\end{abstract}
 \maketitle
 
 \section{Introduction}
 
In recent years, the study of the influence of mass media on the formation of public opinions and cultural globalization has attracted attention in the field of social dynamics \cite{RMP}.
In many sociophysical models, mass media are treated as external or endogenous fields applied on a system of interacting social agents represented as nodes of a network \cite{Galam,Shibanai,GCT,Global,jass,Candia,Peres,Yamir,Croki,Pineda,NJP}.
Mass media influence has also been studied from the perspective of public awareness in models of epidemic spread \cite{Shang}. 
The study of mass media influence on social systems is connected to the general problem considered in Statistical Physics on the competition between particle-particle interactions in a system, that can lead to collective self-organization, and particle-applied field interactions. 
In this context, it has been found that a sufficiently intense mass media field can promote diversity or disorder  
 \cite{Shibanai,GCT,Global,jass,Candia,Peres}, in contrast to the behavior in Ising-type systems, where the elements of the system always tend to align with the state of the field \cite{Kadanoff}. 
These models have mostly considered the effects of a single mass media acting on a system.  The presence of several mass media has been studied in fewer works,
mainly focused on the media influence on the polarization of opinions on a continuous spectrum \cite{Mc,Quattro,Bhat,Take}. 
Social scientists have indicated that polarization of political opinions induced by  exposure to extremist media
constitutes a threat to democracies \cite{McCoy}. On the other hand, 
competition between multiple mass media is a basic feature of modern democratic societies \cite{Strom}. 

 When several external fields are applied in Ising-type systems, the common expectation is that the stronger field dominates. One may ask: are these typical equilibrium concepts generally valid for other interaction dynamics such as those found in social systems? 
In the present article we investigate the collective behavior of a system of social agents subject to the competition between two or more
mass media fields. 
We study 
under what conditions either of the
two mass media acting with different intensities can impose its message to the majority. We find circumstances that lead to the dominance of the weaker mass media. 
We uncover
the emergence of collective ordering in a state alternative to those of the media,
breaking the symmetry of the system in a direction orthogonal to both media fields.
In addition, we explore the dependence of this phenomenon on the topology of the network of interactions.
We show that the presence of long-range interactions is crucial for the occurrence of alternative collective ordering in a state different from those of either mass media.

Instead of a continuous interval of opinions bounded by extremes, we consider discrete, multidimensional equivalent options for social agents.
We employ Axelrod's rules for the dissemination of culture 
as interaction dynamics \cite{Axel}, a paradigmatic non-equilibrium model of wide interest for physicists \cite{Shibanai,GCT,Global,jass,Candia,Peres,Yamir,Croki,Pineda,NJP,Caste,Vespi,Toral,Kuper,Santis,Gracia,Red,Perrier,Pinto}. In Sec.~II, we introduce the model for two mass media influences acting on a fully connected network of social agents. We define some statistical quantities to characterize the collective behavior of the system as its parameters are varied. We identify four collective phases: 
(i) a majority sharing the state of the stronger mass media; (ii) a majority in the state of the weaker mass media; (iii) a majority possessing a state alternative to either mass media; and (iv) a disordered phase.
In Sec.~III, the role of the network topology on the emergence of the alternative ordering and other collective phases is investigated by varying the range of interactions, from local to global
connections. 
In Sec.~IV we show how the model can be extended to include several mass media by considering the case of three mass media fields acting on the system. We find that these effects persist: the formation of a majority in the state of the weakest field and the ordering of the system in a state different from all fields.
Section~V contains the conclusions of this work.

\section{Model for competition between two mass media in a social system}
We consider a system of $N$ social agents consisting of a fully connected network,
where every agent can interact with any other in the system. We assume that interactions take place
according to the dynamics of Axelrod's model for cultural dissemination. In this model,
the cultural state  of agent $i$ $(i=1,2,\ldots,N)$ is given by the $F$-component vector $c_i=(c_i^1, \ldots,c_i^f, \ldots, c_i^F)$,
where each component $c_i^f$ represents a cultural feature that can take any of $q$ different traits or options in the set $\{0, 1,\ldots, q-1\}$.
A mass media influence can be characterized as an external, fixed cultural vector that acts as a global field \cite{GCT,Global,jass}. 
We consider two distinct, external mass media fields acting on the system, denoted respectively by $X=(x^1,\ldots,x^f,\ldots,x^F)$ and $Y=(y^1,\ldots,y^f,\ldots,y^F)$, where 
$x^f$  and $y^f$ are fixed cultural features chosen in the set $\{0,1,\dots,q-1\}$, such as $x^f \neq y^f$ for every $f$.
Here we fix the number of cultural features
to the value $F=10$.

We express the relative strength of the two fields by the parameter $R \in [0,1]$,  such as
 $R$ expresses the strength of $X$ and $(1-R)$ is the strength of $Y$. 
We characterize the intensity of the influence of mass media on the system by the parameter $B \in [0, 1]$, such that at any given time, a randomly selected agent can interact with the mass media with probability $B$ and with any other agent in the system with probability $1-B$, according to the rules of Axelrod's cultural model.
 Then, the product $BR$ represents the
probability for the interaction between an agent and the $X$ field,
and $B(1-R)$ expresses the probability for the interaction between an agent and the $Y$ field.
  At any given time, a randomly selected agent can interact
either with any other agent in the system or with one mass media field, according to the rules
of Axelrod's cultural model.

The states $c_i$ are initially assigned at random with a uniform distribution. 
Then, for given values of parameters $B$ and $R$, 
we define the dynamics of the system by the following iterative algorithm:

\begin{enumerate}
  \item Select at random an \textit{active agent} $i$.
  \item Select the \textit{source of interaction}: (i) active agent $i$ interacts with field $X$ with probability $BR$, (ii) agent $i$ interacts with field $Y$ with probability $B(1-R)$, (iii) agent $i$
    interacts with probability $(1-B)$ with another agent $j$ taken at random in the system.
  \item Calculate the overlap between the active agent $i$ and the source of interaction, 
  defined as the number of shared components between their respective vector states, 
  $d(i,s) = \sum_{f=1}^{F} \delta_{x_i^f,s^f}$,
  where $s^f=x^f$ if the source of interaction is the field $X$, 
   $s^f=y^f$ if the source of interaction is the field $Y$,
  or $s^f=c_j^f$ if $i$ interacts with an agent $j$.
  We use the delta Kronecker function: $\delta_{x,y} = 1$, if $x=y$; $\delta_{x,y} = 0$, if $x\neq y$.
 \item  If $0<d(i,s)<F$,  
  active agent $i$ interacts with probability $d(i,s)/F$ with the selected source of interaction as follows: choose $h$ at random such that $c_i^h \neq s^h$ and set $c_i^h = s^h$.
If $d(i,s) = 0$ or $d(i,s) = F$, the state of agent $i$ does not change.
\end{enumerate}

The interaction dynamics of Axelrod's model on a finite network leads to
an asymptotic configuration characterized by the formation of domains of different sizes.
A domain consists of set of connected agents that share the same state. A homogeneous or ordered collective
state in the system corresponds to a single domain with overlap $d(i, j) = F$, $\forall i, j$.
There are $q^F$ distinct 
realizations for this ordered state. On the other hand, 
an inhomogeneous or disordered collective state possesses several domains. 
In order to investigate the collective behavior arising in the system under the influence of two mass media fields ($B>0, R>0$), we consider
the following statistical quantities calculated in the asymptotic state:

\begin{enumerate}[i)]
\item $S_r$, the average normalized size (divided by $N$) of the domain ranked by the index 
$r$ in the system:  $r=1$ corresponding to the
largest domain, $r=2$ indicates the second largest domain, and so on;
\item $S_x$, the average normalized
size of the domain possessing the state of the field $X$;
\item $S_y$, the average normalized
size of the domain sharing the state of the field $Y$;
\item  $S_{\mbox{\tiny II}}$, the average normalized
size of the largest domain in a state different from those of $X$ and $Y$.
\end{enumerate}

In the absence of fields ($B=0$),  the system reaches
an ordered collective state characterized by $S_1 =1$ for values $q < q_c$, and a disordered state  with $S_1 \to 0$ for $q > q_c$, where $q_c$
is a critical point that depends on the network topology \cite{Caste,Vespi,Toral}. In a fully connected network, the value $q_c$ scales with the system size as $q_c\sim N$ \cite{Red}.

\begin{figure}[h]
\begin{center}
\includegraphics[scale=0.5]{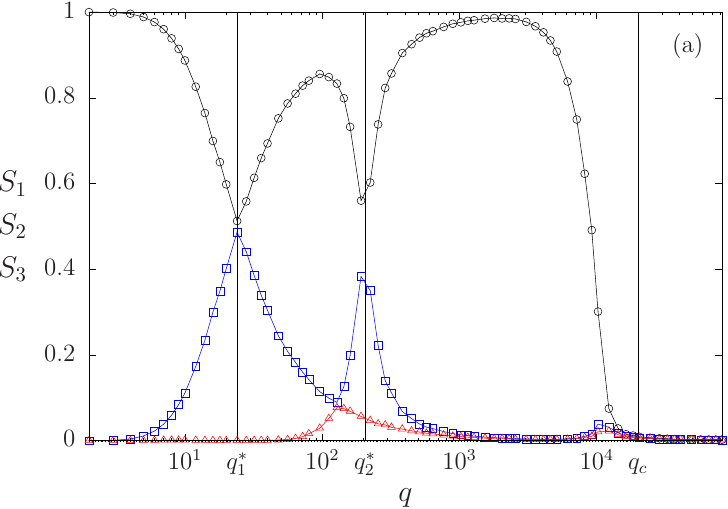} \\
\includegraphics[scale=0.5]{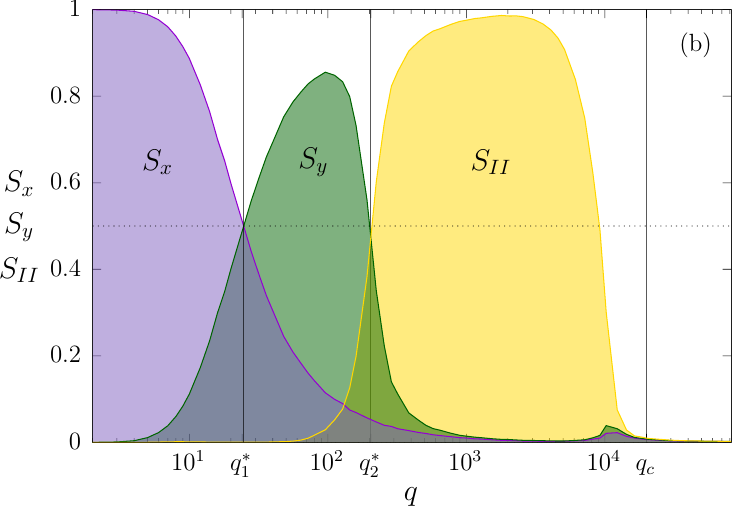}  \\
\includegraphics[scale=0.5]{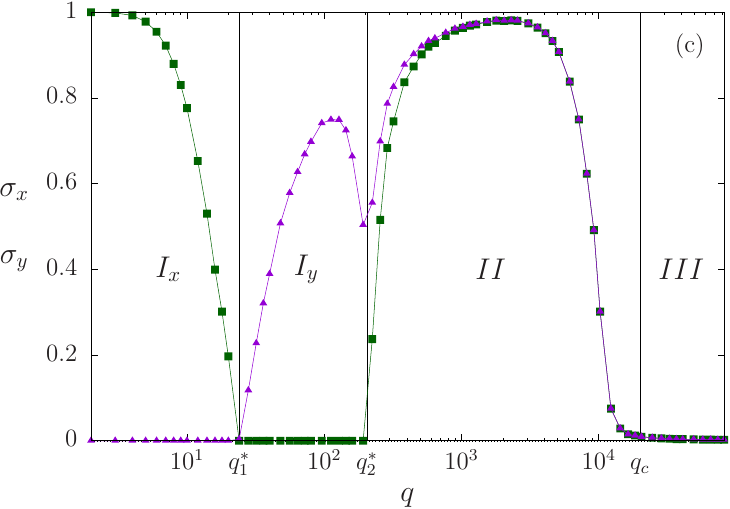} 
\end{center}
\caption{Order parameters for the system subject to two mass media fields $X$ and $Y$ as functions of $q$ (log scale). Fixed parameter values are $B=0.2$, $R=0.92$, and system size $N=2500$.
(a) Average normalized sizes $S_1$ (black line, circles), $S_2$ (blue line, squares), and $S_3$ (red line, triangles) versus $q$ (log scale).
(b)  $S_x$ (magenta line), $S_y$  (green line), and  $S_\text{\tiny{II}}=1-S_x-S_y$  (yellow line) as functions of $q$ (log scale). The areas under the curves are  colored accordingly to illustrate overlapping regions. The horizontal dotted line corresponds to the fraction value $0.5$.
(c) Quantities $\sigma_x$ (magenta line, solid triangles) and $\sigma_y$ (green line, solid squares) as functions of $q$ (log scale). The regions for collective phases $I_x$, $I_y$, II, and III are indicated on the $q$ axis.
  The vertical lines signal the values $q_1^*$, $q_2^*$, and $q_c$ on each panel. Each data point is obtained from $50$ realizations of initial conditions.}
\label{f1}
\end{figure}

When two mass media fields are present, each field competes with the other to establish its state on the system. In addition,
 the spontaneous order emerging in the system due to the agent--agent interactions for values $q<q_c$  competes with the ordering being imposed by the fields. 
 These processes can be ascertained in Fig.~\ref{f1}(a) shows the average normalized sizes  $S_1$, $S_2$, and $S_3$ calculated as functions of the number of options $q$ when the two mass media  are applied. 
In Fig.~\ref{f1}(a), parameter $R>0.5$ such that the mass media field $X$ is more intense than the field $Y$. 
The quantity $S_1$ exhibits two sharp local minima at values $q_1^*$ and $q_2^*$, where $q_1^*<q_2^*<q_c$.
We observe that the local minima of $S_1$ correspond to local maximum values of $S_2$. 
We find that $S_1 +S_2 +S_3 \approx 1$ for $q < q_c$. Therefore, three domains occupy almost the entire system for $q < q_c$.
For $q>q_c$, the system reaches a disordered state where $S_1 \to 0$.

To investigate the nature of the minima in $S_1$ and the maxima in $S_2$, we plot in Fig.~\ref{f1}(b) the domain sizes $S_x$, $S_y$, and $S_{\text{\tiny II}}=1-S_x-S_y$ as functions of $q$.
In the region $q < q_1^*$, 
the domain of size $S_x$ that shares the state of the 
stronger field $X$ constitutes a majority ($S_x>0.5$) and it is equal to $S_1$. 
The remaining agents in the system have states non-overlapping with $X$
but they can interact with the weaker field $Y$, giving rise to the second largest domain in the state of $Y$, i.e.,  $S_2=S_y$. As $q$ increases, there are more agents in states orthogonal to $X$ but still able to interact with $Y$.  Thus, $S_x$ decreases while $S_y$ grows in size until they become equal, $S_y=S_x=0.5$, at the value $q=q_1^*$, where the roles of the fields are interchanged: the majority  acquires the state of $Y$ and the second largest domain of size $S_2$ lies in the state of $X$.
Then, $S_y$ becomes a majority ($S_y>0.5$) for values $q_1^* < q < q_2^*$.
As $q$ is further increased, there are many states in the system that are orthogonal to both fields.  Then, $S_y$ decreases while $S_{II}$ increases until $S_y=S_x=0.5$ at the value $q=q_2^*$.
For values $q_2^*<q<q_c$, more states in the system are orthogonal to both $X$ and $Y$;  the agent-agent interactions prevail over the agent-field interactions and the system spontaneously organizes into one of those orthogonal states. The largest domain $S_1$ and the majority emerging in the system 
corresponds to an ordered, alternative state non-overlapping with either mass media field $X$ or $Y$.
For values $q > q_c$, the system settles into a disordered state.  

To characterize the collective behavior of the system, we define the following quantities as order parameters, 
\begin{eqnarray}
 \sigma_x &=& S_1-S_x , \\
 \sigma_y &=& S_1-S_y .
\end{eqnarray}

We show the quantities $\sigma_x$ and  $\sigma_y$ as functions of $q$ in Fig.~\ref{f1}(c). 
Four regions on the parameter $q$ can be distinguished in Fig.~\ref{f1}(c), corresponding to different collective states:
\begin{enumerate}[1)] 
 \item  A region $q<q_1^*$ denoted by $I_x$, characterized by $\sigma_x=0$ and $\sigma_y>0$, where the more intense field $X$ imposes its state to the majority.
 \item  A region $q_1^* < q < q_2^*$ labeled $I_y$, characterized by $\sigma_x>0$ and $\sigma_y=0$, where the weaker field $Y$ imposes its state to the majority.
 \item A region $q_2^*<q<q_c$ indicated as II, characterized by $\sigma_x>0$ and $\sigma_y>0$, where a majority emerges in a state different from those of $X$ and $Y$.
 \item A disordered phase for $q>q_c$ denoted by III, characterized by $S_1 \to 0$ ($\sigma_x=0$ and $\sigma_y=0$). The value $q_c$ is found to be independent on $R$.
\end{enumerate}

\begin{figure}[h]
	\includegraphics[scale=0.5]{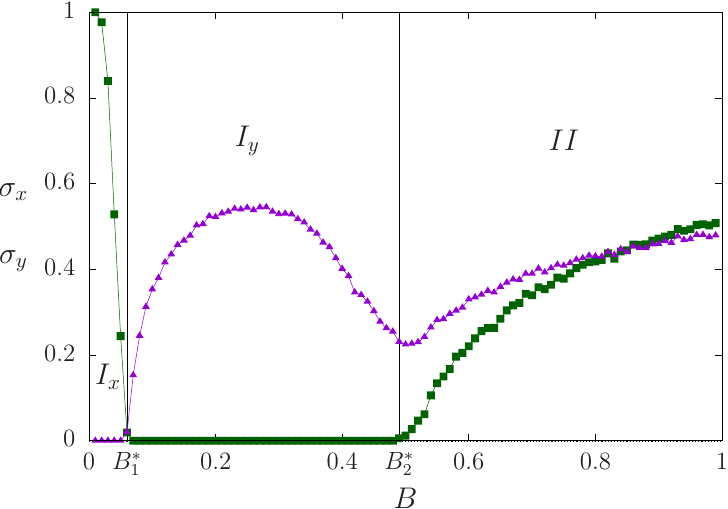} 
\caption{Order parameters $\sigma_x$ (magenta line, solid triangles) and $\sigma_y$ (green line, solid squares) as functions of $B$. Vertical lines signal the values $B_1^*$ and $B_2^*$. Labels indicate phases $I_x$ ($S_1$ has the state $X$), $I_y$ ($S_1$ possesses the state $Y$), and II ($S_1$ has an alternative state). 
Fixed parameters are $q=50$, $R=0.92$, system size $N=2500$. Each data point is obtained from  $50$ realizations  of initial conditions.}
\label{f2}
\end{figure}

In Fig.~\ref{f2} we show the dependence of  $\sigma_x$ and $\sigma_y$ on the intensity of the media $B$ for fixed $q$. 
For values of $B$ below some threshold $B_1^*$, the stronger field $X$ imposes its state on the majority, giving rise to phase $I_x$ ($\sigma_x=0$ and $\sigma_y>0$). 
Increasing $B$ means that the stronger field $X$ will more rapidly attract a small fraction of agents that overlap with $X$. 
The remaining larger number of agents are able to interact with $Y$ and among themselves. In this range of $q$ values, the agent-$Y$ interaction wins over the agent-agent interaction because the probability to interact with $Y$ also increases with increasing $B$. 
Then, the weaker field $Y$ drives 
the majority to its state and phase $I_y$ ($\sigma_x>0$ and $\sigma_y=0$) arises on a range of values $B_1*<B<B_2^*$.
When $B$ is further increased, field $Y$ follows the same fate that $X$; it will attract a small fraction of the agents, but the remaining agents are in many states orthogonal to $Y$ (and to $X$). These agents constitute now a majority that interact among themselves to reach a homogeneous state resulting from the absorbing dynamics. This homogeneous state constitutes phase II in a state orthogonal to both $X$ and $Y$.

\begin{figure}[h]
\includegraphics[scale=0.5]{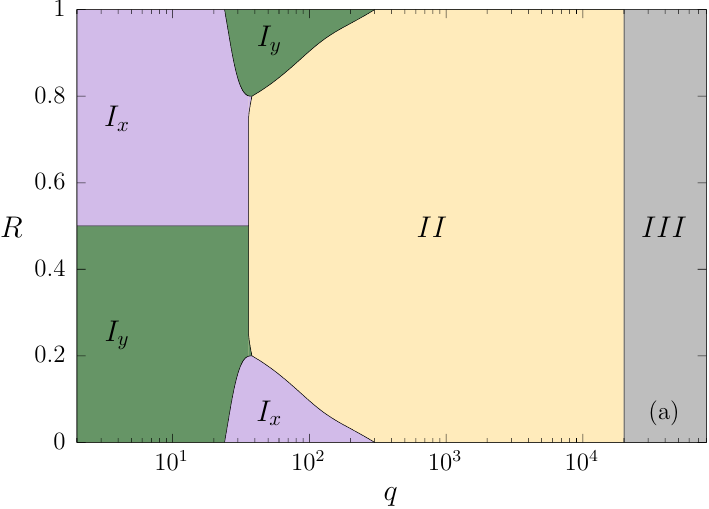} 
\includegraphics[scale=0.5]{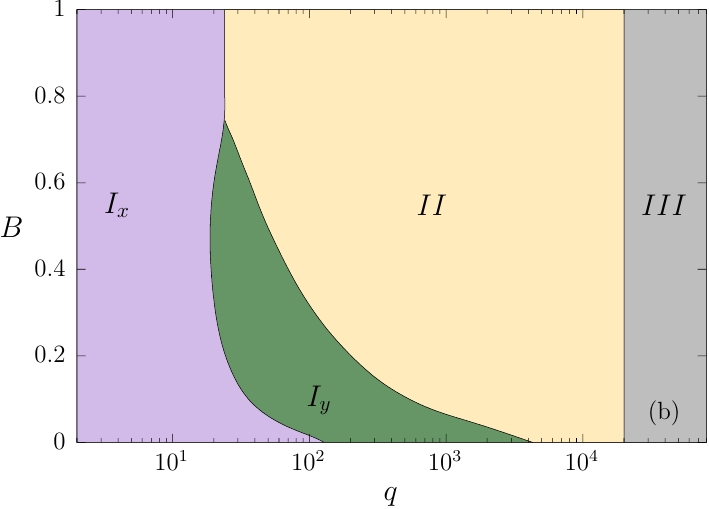}
\caption{a) Phase diagram on the space of parameters $(R,q)$  for fixed $B=0.2$. b) Phase diagram on the space of parameters $(B,q)$  for fixed $R=0.92$. Collective phases are characterized by the quantities $\sigma_x$ and $\sigma_y$. For both phase diagrams, system size is $N=2500$ and each data point is calculated over $10$ realizations of random initial conditions.  Labels and colors indicate the different phases: $I_x$, magenta (the largest domain $S_1$ has the state $X$); $I_y$, green (the largest domain $S_1$ possesses the state $Y$); II, yellow (the largest domain $S_1$ has an alternative state); III, gray (disordered state).}
\label{f3}
\end{figure}

Figure~\ref{f3}(a) shows the phase diagram or collective states on the space of parameters 
$(q,R)$ for the fully connected network under the influence of the mass media fields $X$ and $Y$, with fixed value $B$. 
The field $X$ is stronger than the field $Y$ for $R>0.5$ and the largest domain acquires the state of $X$ on the region denoted by $I_x$.
For parameter values $R<0.5$, the field $Y$ is more intense than $X$. Then, the regions $I_x$ and $I_y$ become interchanged on the plane $(q,R)$ with respect to the line $R=0.5$. 
 The regions $I_x$ and $I_y$ ($I_y$ and $I_x$) on the $q$ axis in Fig.~\ref{f3}(a) reflect the competition between the two mass media fields that ensues when the intensity $B$ is small; i. e., when the agent-field interactions are less likely.

Figure~\ref{f3}(b) shows the phase diagram of the system on the space of parameters
$(q,B)$. Both Fig.~\ref{f3}(a) and Fig.~\ref{f3}(b) indicate that the stronger mass media always imposes its state to the majority for small values of the number of options $q$.
On the other hand, the weaker mass media field can prevail on the system for intermediate values of $q$.
Counterintuitively, neither mass media is effective when the intensity $B$ is large enough; a large domain possessing a state orthogonal to both mass media fields emerges in the system.

\begin{figure}[h]
\includegraphics[scale=0.54]{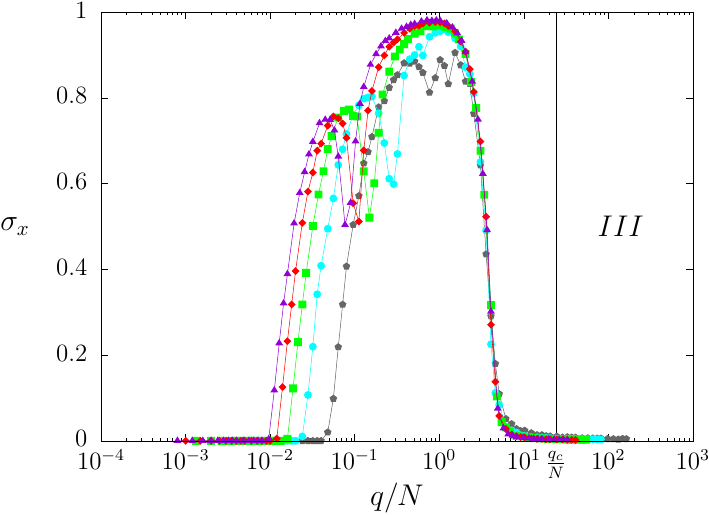} 
\caption{The quantity $\sigma_x$ as a function of $q/N$ for several values of $N$: $N=2500$ (triangles), $N=2000$ (diamonds), $N=1500$ (squares), $N=1000$ (circles), $N=500$ (pentagons).
Fixed parameters: $B=0.2$, $R=0.92$. Each data point is obtained from $50$ realizations of initial conditions.}
\label{f4}
\end{figure}

To explore the behavior of the system for different system sizes $N$, Fig.~\ref{f4}
shows the quantity $\sigma_x$ as a function of $q_c/N$ for different values of $N$. The critical point for the transition to the disordered phase III is constant, indicating that $q_c \sim N$ even in the presence of external fields. In comparison, the values of $q_1^*$ and $q_2^*$ change with increasing $N$. We have verified that phases $I_x$, $I_y$, II, and III continue to form as $N$ increases. 

\section{Influence of the network topology}

To investigate the role of the topology of the network on the collective phases emerging in the system,
we first consider a system consisting of a ring of $N+1$ agents or nodes where each agent is connected to $k$ neighbors, $k/2$ on each side. The ratio $k/N$ characterizes the range of the interactions; nearest-neighbor coupling is described by $k/N \to 0$ ($N$ large), while
the fully connected network studied above corresponds to the value $k/N=1$. As in Sec.~II, the system is subject to the influence of two global mass media fields $X$ and $Y$ possessing relative strength $R$ and acting on the agents with probability $B$.  Then, the probability of
interaction between a selected agent and any of its $k$ neighbors is $1-B$.

\begin{figure}[h]
	\includegraphics[scale=0.5]{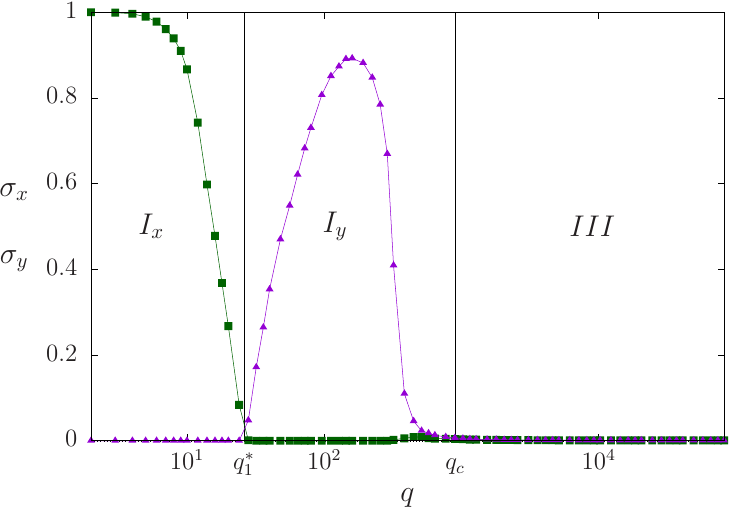} 
	\caption{Quantities $\sigma_x$ (magenta solid triangles), and $\sigma_y$ (green solid squares) as functions of $q$ (log scale) for a ring network with interaction range $k/N=0.028$ subject to two external fields $X$ and $Y$. Fixed parameter values:
	$B=0.2$, $R=0.92$, $k=72$, and system size $N=2501$. Each data point was calculated over $50$ realizations of random initial conditions.}
	\label{f5}
\end{figure}

Figure~\ref{f5} shows the quantities $\sigma_x$ 
and $\sigma_y$ as functions of $q$ for the ring network with a fixed a short range of interaction $k/N$. Ordered phases $I_x$, $I_y$, and disordered phase III take place, but the alternative phase II  does not emerge in the system. In comparison,  phase II occurs in the fully connected network for a range of values of $q$, as shown in Fig.~\ref{f1}(c).

\begin{figure}[h]
	\includegraphics[scale=0.55]{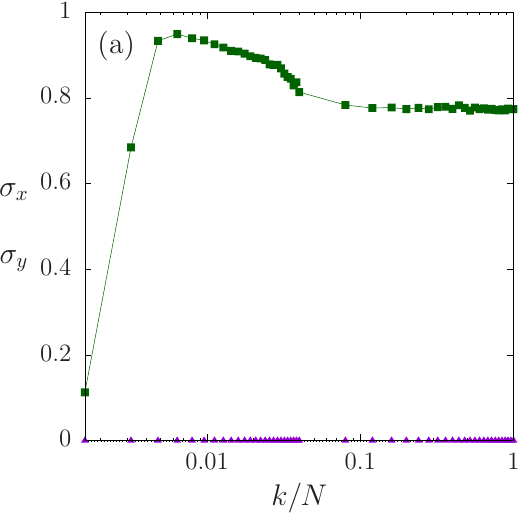} 
	\includegraphics[scale=0.55]{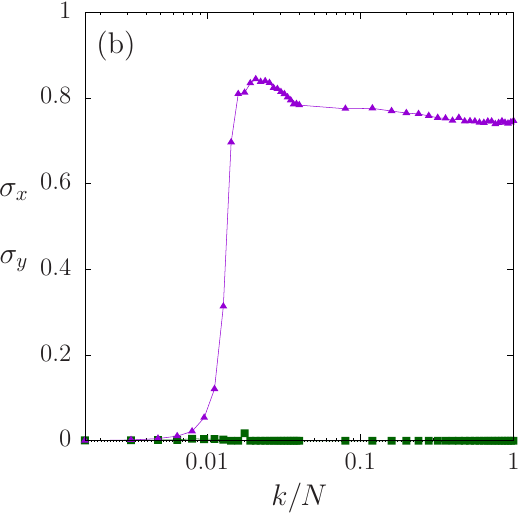}
\includegraphics[scale=0.55]{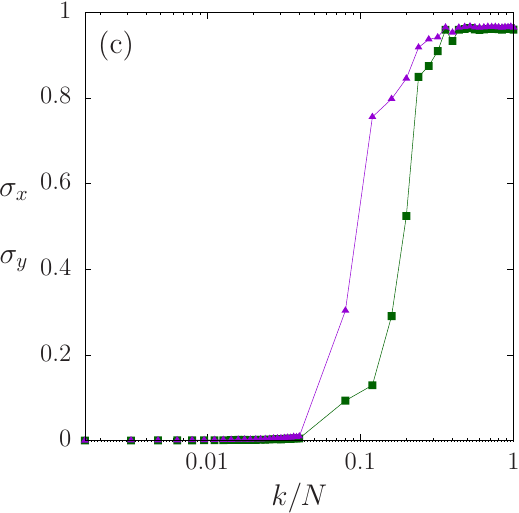}
	\caption{Quantities $\sigma_x$ (magenta solid triangles), and 
$\sigma_y$ (green solid squares) as functions of $k/N$ (log scale) in a 
	 ring network subject to external fields $X$ and $Y$, for different values of $q$. Fixed parameters are $B=0.2$, $R=0.92$, and system size $N=2501$. 
Each data point was calculated over $50$ realizations of random initial conditions.
 a) $q=10$. b) $q=100$. c) $q=1000$.}
		\label{F6}
\end{figure}

In Fig.~\ref{F6} we plot the quantities $\sigma_x$ 
and $\sigma_y$  for the ring network as functions of the ratio $k/N$, for three different values of the number of options $q$. 
Figure~\ref{F6}(a) shows that, when the number of options is small, the ordered phase $I_x$ induced by the stronger field $X$ occurs for all ranges of interaction, from nearest-neighbor coupling to all-to-all coupling. Figure~\ref{F6}(b) shows that the ordered phase $I_y$ produced by the weaker field arises for $k/N>0.01$; in particular, this phase does not appear for nearest-neighbor coupling.
Similarly, Fig.~\ref{F6}(c) reveals that the alternative phase II ($\sigma_x>0$, $\sigma_y>0$) emerges for a longer range of interaction, above the critical value  $k/N=0.04$, and therefore it could not be observed in Fig.~\ref{f5}.
Thus, both phase $I_y$ and alternative phase II arise when a critical range of interaction is reached for each phase.

To explore the influence of randomness of the connections, we consider a small-world network of size $N=2500$  and degree $k$, with a random rewiring probability $p$ following the Watts-Strogatz algorithm \cite{Watts}. A probability $p=0$ corresponds to a ring of $N$ nodes where each node is connected to $k$ neighbors, while $p=1$  describes a fully random network. The fully connected network occurs when $k/N =1$.

\begin{figure}[h]
	\includegraphics[scale=0.5]{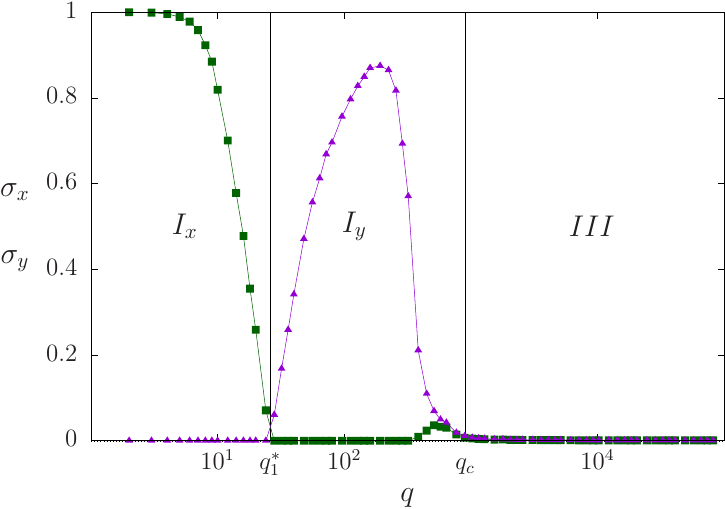} 
	\caption{Quantities $\sigma_x$ (magenta solid triangles) and $\sigma_y$ (green solid squares) as functions of $q$ (log scale) for a small-world network with size $N=2500$, $k=72$, and rewiring probability $p=0.02$, subject to two external fields $X$ and $Y$. Fixed parameter values
	$B=0.2$, $R=0.92$. Each data point was calculated over $50$ realizations of random initial conditions.}
	\label{f7}
\end{figure}

Figure~\ref{f7} shows the quantities $\sigma_x$ 
and $\sigma_y$ as functions of $q$ for the small-world network with $p=0.02$ and $k=72$ subject to two mass media fields $X$ and $Y$. The range of interaction $k/N=0.028$ is equal to that of the ring network in Fig.~\ref{f5}. Again, we observe ordered phases $I_x$, $I_y$, and the disordered phase III, but the alternative phase II does not appear, suggesting that it may require a longer range of interaction.

\begin{figure}[h]
	\includegraphics[scale=0.56]{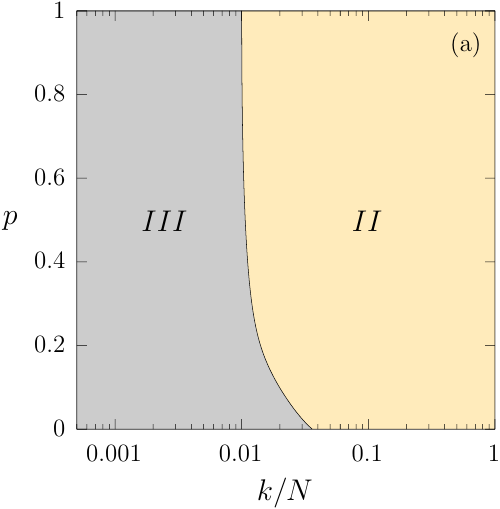}
\includegraphics[scale=0.55]{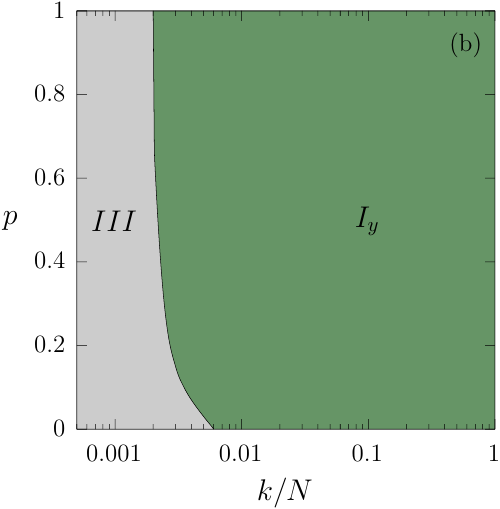}
	\caption{Collective behavior on the space of parameters $(p, k/N)$ for a small-world network ($k/N$ in log scale). Fixed values are $B=0.2$, $R=0.92$, system size $N=2500$. 
	a) Region where alternative phase II emerges (yellow color, labeled as II) for fixed $q=1000$.
	b)  Phase $I_y$ (green color, labeled $I_y$) for $q=100$. 
Each data point was calculated over $50$ realizations of random initial conditions.}
    \label{f8}
\end{figure}

Figure~\ref{f8} displays the collective phases, characterized by the quantities $\sigma_x$ and $\sigma_y$, on the space of parameters
$(p, k/N)$ for a Watts-Strogatz small-world network and different values of $q$.
Figure~\ref{f8}(a) shows a critical boundary on this space of parameters that separates the transition from 
a disordered phase III, where $\sigma_x=0$ and $\sigma_y=0$, to an alternative phase II characterized by $\sigma_x>0$ and $\sigma_y>0$.
For increasing values of the rewiring probability $p$, the transition happens for a constant critical value of about $k/N= 0.01$.  Similarly,
Fig.~\ref{f8}(b) shows that the region where phase $I_y$, characterized by $\sigma_x>0$ and $\sigma_y=0$, arises on this space  for values of $k/N$ above a smaller critical value.
Thus, random connectivity is not the network property mostly contributing to the rise of the alternative phase II or phase $I_y$; rather these effects appear
when a critical range of interactions $k/N$ is reached in each case.

Our results indicate that the alternative phase II and phase $I_y$ take place when a sufficiently long range of interactions or a critical degree of connections is present in a network. This critical degree is a small fraction of the size of the system. In particular, all-to-all interactions are not necessary for the growth of a majority group in a state different from those of the two mass media influences acting on the network.

\section{Multiple mass media competition}
The model can be extended to include several mass media fields acting on the system described by a fully connected network. For example, consider three mass media described by the vectors
$X=(x^1,\ldots,x^f,\ldots,x^F)$, $Y=(y^1,\ldots,y^f,\ldots,y^F)$ and
$Z=(z^1,\ldots,z^f,\ldots,z^F)$ such that $X$, $Y$ and $Z$ are orthogonal, i. e.,
$x^f\neq y^f \neq z^f$, $\forall f$. We define the relative strength of the fields $X$, $Y$ and $Z$ by the parameters $R_x$, $R_y$, and $R_z$, respectively, such that $R_x+R_y+R_z=1$. Again, the parameter $B$ characterizes the intensity of the mass media influence. Then, $BR_x$, $BR_y$, and $B(1-R_x-R_y)$ represent the probability for an agent to interact with the fields $X$, $Y$, and $Z$, respectively, while $1-B$ is the probability for agent-agent interaction.

\begin{figure}[h]
\begin{center}
\includegraphics[scale=0.5]{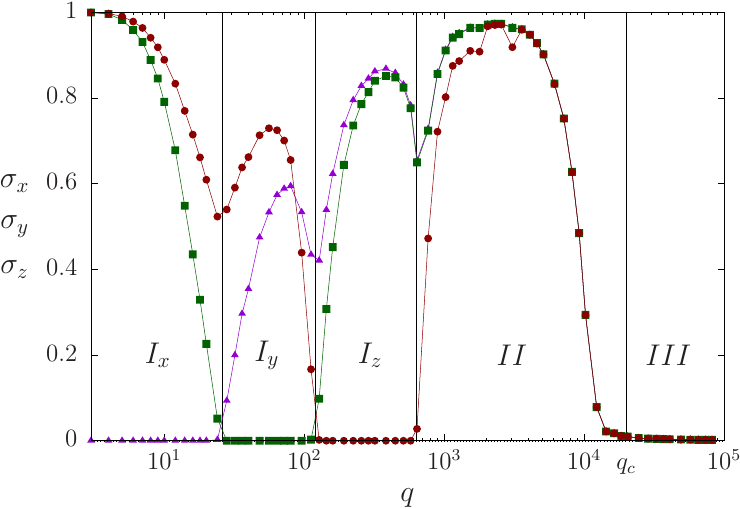} 
\end{center}
\caption{Quantities $\sigma_x$ (magenta line, solid triangles), $\sigma_y$ (green line, solid squares), and $\sigma_z$ (red line, solid circles) as functions of $q$ (log scale). 
Fixed parameter values are $B=0.2$, $R_x=0.9$, $R_y=0.08$, and $R_z=0.02$
System size $N=2500$.
The regions for collective phases $I_x$, $I_y$, $I_z$, II, and III are indicated on the $q$ axis.
  Vertical lines delimit the different phases on the $q$ axis. Each data point is obtained from $50$ realizations of initial conditions.}
\label{f9}
\end{figure}

Figure~\ref{f9} shows the quantities $\sigma_x =S_1-S_x$,  $\sigma_y= S_1-S_y$, and  
$\sigma_z= S_1-S_z$ as functions of $q$. The relative strengths of the fields are
$R_x >R_y>R_z$. Five regions on the parameter $q$ can be distinguished in Fig.~\ref{f9}. They correspond to:
(i) phase $I_x$ where field $X$ imposes its state to the majority, characterized by $\sigma_x=0$, $\sigma_y>0$, $\sigma_z>0$;
(ii) phase $I_y$ having a majority in the state of $Y$, characterized by $\sigma_x>0$,  $\sigma_y=0$, $\sigma_z>0$;
(iii) phase $I_z$ where a majority shares the state of $Z$, characterized by $\sigma_x>0$,  $\sigma_y>0$, $\sigma_z=0$;
(iv) alternative phase II, characterized by $\sigma_x>0$,  $\sigma_y>0$, $\sigma_z>0$, where a majority emerges in a state different from those of $X$, $Y$, and $Z$;
(v) disordered phase for $q>q_c$ denoted by III, characterized by $\sigma_x=0$, $\sigma_y=0$, $\sigma_y=0$ $(S_1 \to 0)$. 

The stronger field $X$ drives the majority to its state for small values of $q$. Then, each field $Y$ and $Z$, in decreasing value of strength, imposes its state on the majority on successive intervals of increasing values of $q$. For large values of $q <q_c$ a majority emerges in an alternative state non-overlapping with any mass media field. Again, we find two non-trivial effects: (i) the weakest mass media message can convince the majority on a range of values of $q$, and (ii) the system can spontaneously order against all applied fields.

\section{Conclusions}
We have investigated the collective behavior of a network of interacting social agents subject to the competition between two mass media sources. 
Besides the natural interest of this question for social and political sciences, 
the competition between self-organization and multiple external fields in non-equilibrium systems, as well as the influence of network topology on the collective behavior, is a relevant problem in complex systems. 
Our model is based on Axelrod's rules for cultural dynamics that allow for non-interacting states between the agents, and between the agents and the fields.
This type of interaction is common in social and biological systems where there is often some bound or condition for the occurrence of interaction.

By studying this model on a fully connected network, we have characterized four collective phases on the space of parameters of the system: 
(i) a majority in the state of the stronger mass media; (ii) a majority sharing the state of the weaker mass media; (iii) a large domain possessing a state alternative to either mass media; and (iv) a disordered phase.
We have found that
the stronger mass media always imposes its state on the majority for small values of the number of options $q$.
On the other hand, the weaker mass media field can prevail on the system for intermediate values of $q$.
Counterintuitively, neither mass media is effective when their intensity $B$ is large enough; a large domain possessing a state orthogonal to both mass media fields emerges in the system.
The rise of this alternative group reflects the tendency towards the spontaneous order associated to the agent-agent interactions when the number of options is large enough.  

We have studied the influence of the network topology, mainly the range and the randomness of the connections, on the collective behavior of the system.
By considering a ring network with a varying range of interactions, we have shown that the stronger mass media field imposes its state on the system for all ranges of interactions.
In contrast, the convincing power of the weaker mass media depends on the
presence of a minimum number of long-range connections in the network.
Similarly, 
the rise of alternative ordering 
is related to the existence of a critical number of connections in the 
network. We have also considered a small world network with varying range of interactions and rewiring probability. We found that 
randomness of the connections is not the main
network property contributing to 
the prevalence of the weaker mass media field and the spontaneous ordering against the global fields; 
rather, these effects appear when a critical range of interactions is reached in each case. 
Thus, all-to-all interactions are not essential for observing these phenomena. 

We have shown how the model can be extended to include several mass media by considering three mass media fields acting on the system. Nontrivial collective behaviors persist with multiple fields for some parameter values: the weakest mass media can convince the majority and the system can order against all applied fields.

The success of a weaker mass media message and the emergence of an alternative group not aligned with any mass media should be apparent in experiments measuring media influence, product adoption, or marketing in social networks.
Our results suggest that the emergence of an ordered phase with a state different from those of multiple applied external fields should arise in other non-equilibrium systems possessing non-interacting states. Future research problems should include the competition between overlapping fields, competition between external and endogenous mass media fields such as cultural trends, competition of external mass media versus influencers in a social network, the propagation of mass media messages varying in time and space, and 
the consideration of diverse interaction rules.

 \section*{Acknowledgment}
 This research was supported by Corporaci\'on Ecuatoriana para el Desarrollo de la
 Investigaci\'on y Academia (CEDIA) through project CEPRA XVI-2022-09 ``Aplicaciones en Sociof\'isica".

\end{document}